\documentclass[article]{IEEEtran}
\usepackage{graphicx}
\usepackage{grffile}
\usepackage{setspace}
\usepackage{longtable}
\usepackage{url}
\usepackage{cite}
\usepackage{multirow}
\usepackage{tabularx}
\usepackage{amsmath}
\usepackage{subcaption}
\usepackage{algorithm,algpseudocode}

\usepackage[usenames, dvipsnames]{color}

\title{Large Scale Deterministic Networking:\\ A Simulation Evaluation
\thanks{Please direct correspondence to M.~Reisslein (reisslein@asu.edu).}}

\author{Ahmed Nasrallah, Venkatraman Balasubramanian, Akhilesh Thyagaturu, Martin Reisslein, and Hesham ElBakoury}

\begin{document}
\maketitle

\begin{abstract}
The use of Ethernet switched networks usually involves best effort
service. A recent effort by the IEEE 802.1/3 TSN group has sought to standardize
the Ethernet data-link protocol such that it operates on a
deterministic service in addition to the best effort service targeting
Operational Technology applications, e.g., industrial control
systems. This paper investigates the Cyclic Queueing and Forwarding
(CQF) and Paternoster scheduling protocols in a typical industrial
control loop with varying propagation delays emulating large scale
networks. Our main findings for CQF and Paternoster are that CQF has an
advantage towards real-time streams with hard-deadlines whilst
Paternoster is for streams with more relaxed deadlines but can operate
without time synchronization.
\end{abstract}


\section{Introduction}

\subsection{Motivation}
\begin{figure} [t!] \centering
	\includegraphics[width=3.5in]{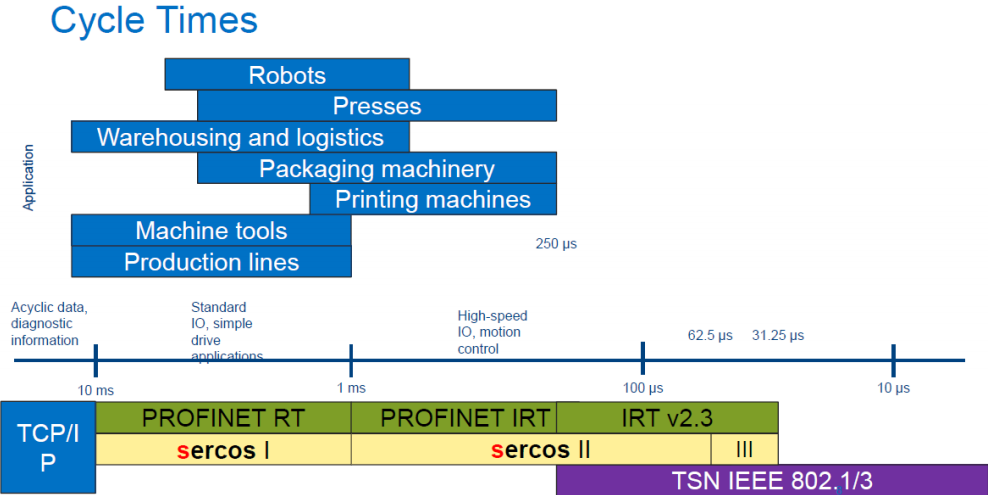}
	\caption{Industrial QoS between different protocols related to OT applications.}
	\label{OT-pic}
\end{figure}

The open access and ubiquitous use of Ethernet switched networking technology is propelling the use of full-duplex Ethernet standards in LANs and WANs for a variety of real-time and traditional background applications on converged Ethernet switches and links. The use of Ethernet in industrial environments provides increased bandwidth and better interoperability among other benefits. While the idea of using Ethernet devices in Operational Technology (OT), i.e., automotive, avionics, and industrial control systems, is not new (see Fig.~\ref{OT-pic}), the IEEE 802.1Q, Time-Sensitive Networking (TSN), task force recently released a set of standards that augment standard Ethernet switches providing determinism and low latency communication ideal for OT applications.

A key question that needs defining is what constitutes a deterministic system or determinacy in the context of networking and communication? We can establish that it does not mean increased throughput or reduced latency. We conclude that a deterministic system is a system in which \textit{no} randomness is involved and therefore can be modeled or characterized to produce the same output from the same starting conditions (i.e., initial state).

In this report, we implement and utilize Cyclic Queuing and Forwarding (CQF), and the Paternoster scheduling mechanism on a standard industrial control closed-loop unidirectional ring topology. Furthermore, we study and analyze the scheduling mechanism's efficacy for different propagation delays and traffic intensity emulating large-scale networks with both sporadic and periodic traffic. The main goal is to ensure the deterministic attributes governed by the scheduling mechanism used in ensuring proper TSN QoS.

\subsection{Related Work} \label{tsn:rel:sec}
Groundwork on CQF, which was also previously known as Peristaltic shaper, was conducted by Thangamuthu et al.~\cite{Thangamuthu2015}.  Moreover, Thiele et al.~\cite{thiele2015timeaware} have conducted a theoretical analysis of the blocking factors for CQF and TAS.

Zhou et al.~\cite{zhouinsight,zho2018ana} have conducted a simulation study on Paternoster, but only for one-hop transmission (they did not consider a full multi-hop network).

In \cite{Ata10.1007/978-3-030-22999-3_44} authors model a routing problem in Time sensitive network as an ILP  in time sensitive networks.

In \cite{pahlevan2019heuristic} authors propose a joint optimization problem of routing and scheduling in one step.

In \cite{heilmann2019size} authors propose a bandwidth optimization based queuing technique.

\subsection{Contributions}
We make the following contributions:

\begin{itemize}
\item[i)] We implement both standard CQF and Paternoster scheduling models.
\item[ii)] We comprehensively evaluate and analyze the two models for both sporadic and periodic sources with cross-interference of BE traffic and varying propagation delays emulating large-scale networks.
\item[iii)] We elucidate recommendations and limitations of each model according to the results.
\end{itemize}

\subsection{Organization}
This article is organized as follows. Section~\ref{tsn:back:sec} provides the necessary background information to understand the mechanisms of CQF and Paternoster. Section~\ref{tsn:sim:setup} describes and illustrates the simulation environment, network/traffic model, and shows the results collected and metrics involved in analyzing the scheduling mechanisms. Finally, Section~\ref{concl:sec} concludes the document.

\section{Background: IEEE 802.1 Time Sensitive Networking} \label{tsn:back:sec}
This section provides a brief background overview on TSN standardization, specifically CQF and Paternoster. TSN is a suite of standards aimed at applying deterministic behavior to the traditional best-effort Ethernet standards. Due to the scope of this project, we refer to the survey~\cite{nas2019ult} for further reading on TSN standardization and research areas.

\subsection{CQF}

\begin{figure} [t!]
	\centering
	\includegraphics[width=3.5in]{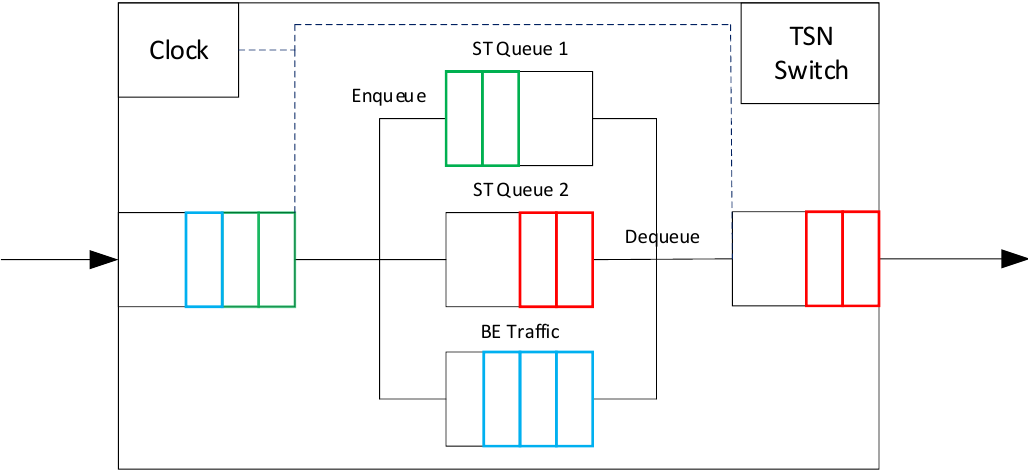}
	\caption{Simplified CQF Mechanism in TSN switch with two ST queues}
	\label{cqf}
\end{figure}

The published IEEE 802.1Qch (CQF)~\cite{IEEE8021Qch} standard proposes to coordinate enqueue/dequeue operations within a switch in a cyclic fashion. Fig~\ref{cqf} shows a simplified illustration (or snapshot) of the CQF mechanism. The CQF cyclic operation results in an easily calculable latency bound governed by the chosen Cycle Time and the number of end-to-end hops between communicating parties. In CQF, time is divided into slots or intervals. For a given traffic class, two queues are used to enable the cyclic property. Frames arriving in interval $x$ will be transmitted in interval $x+1$. Similarly, frames arriving in interval $x+1$ are transmitted in interval $x+2$, and so on.  The maximum and minimum frame delay bounds in CQF with $H$ and $CT$ representing the number of hops and cycle time duration, respectively, are
\begin{eqnarray}	\label{eq1}
D_{Max} &=& (H+1) \times CT  \\
D_{Min} &=& (H-1) \times CT.  	\label{eq2}
\end{eqnarray}
Two queues are used to handle enqueue and dequeue operations in separate time intervals. For example, frames arriving in even intervals will be enqueued in one queue, while the frames that were enqueued during the previous interval will be transmitted from the other queue. In CQF, a frame sent by an upstream switch in cycle $x$ must be received by the downstream at cycle $x$, i.e., the propagation delay must be less than the selected cycle time. Therefore, the cycle time is constrained by the link distance (network scale in general). Essentially, the smaller the network size, the easier it is to guarantee the TSN QoS by CQF. Additionally, CQF has a few challenges that limit its viability, such as $i)$ accurately determining the appropriate cycle time, and $ii)$ cycle duration misalignment where due to processing and transmission delays, a frame can be received in the wrong cycle (i.e., be placed in the wrong outbound queue).

\subsection{Paternoster}

\begin{figure} [t!]
	\centering
	\includegraphics[width=3.5in]{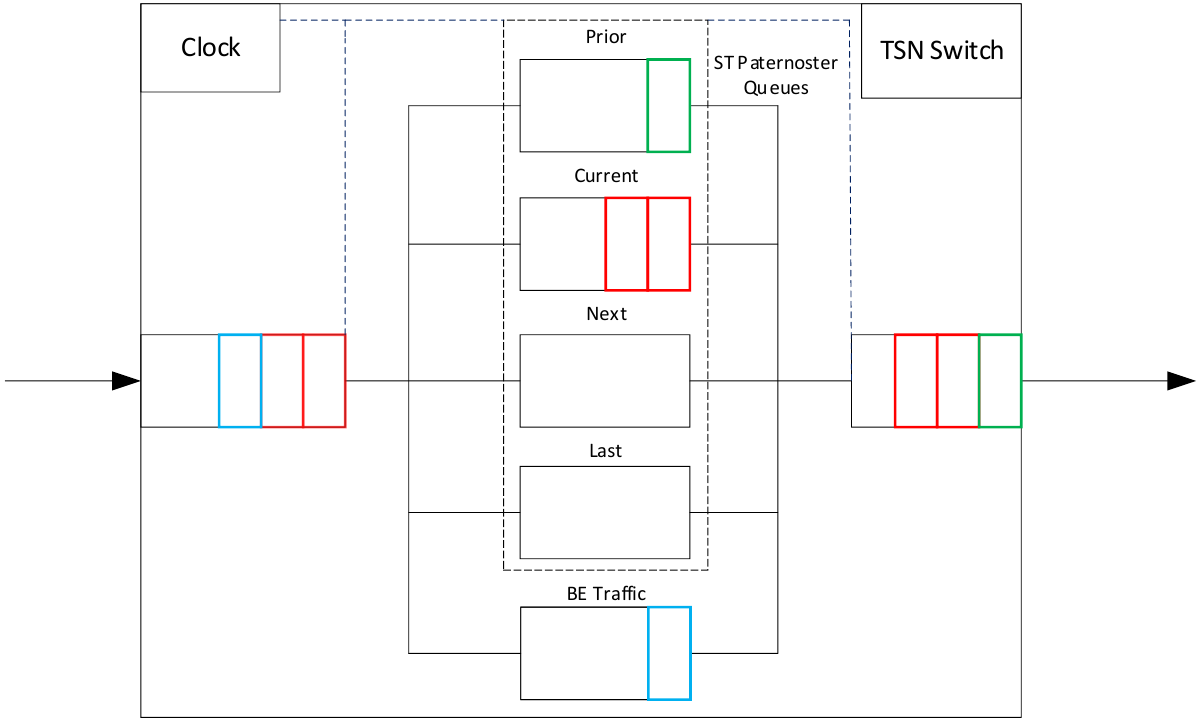}
	\caption{Simplified Paternoster Mechanism in TSN switch}
	\label{pat}
\end{figure}

The Paternoster algorithm is a proposed enhancement by Mike Seaman~\cite{seaman2019pat} to standard CQF. Fig~\ref{pat} shows a simplified illustration (or snapshot) of the Paternoster mechanism. Paternoster provides bounded latencies and lossless service for flows that are successfully registered across the network without a time synchronization requirement. For each egress port, the Paternoster protocol defines a counter for stream reservation and four output queues (\textit{prior}, \textit{current}, \textit{next}, \textit{last}), whereby all switches under Paternoster operate under an \textit{epoch} timescale where the stand/end of the \textit{epochs} are not synchronized with other switches. In each \textit{epoch} window, frames in the \textit{prior} queue are transmitted first until all frames are transmitted. Once the \textit{prior} queue is depleted, the \textit{current} queue is selected for transmission until the end of the current epoch. While frames are being transmitted from the \textit{prior} and \textit{current} queues, received frames are enqueued in the \textit{current} queue until the bandwidth capacity is reached for the current \textit{epoch}. Any additional frames are enqueued in the \textit{next} and \textit{last} queues in a similar manner, i.e., until the reservation capacity for the current epoch is reached while additional frames are dropped if the \textit{last} queue is completely reserved for the current \textit{epoch}. Note that all ST traffic streams are given guaranteed bandwidth, while BE traffic is given the leftover bandwidth. When a new epoch starts, the previous \textit{current} queue operates as the \textit{prior} queue while the \textit{next} and \textit{last} queues become the \textit{current} and \textit{next} queues, respectively. The previous \textit{prior} queue (which should be empty, and if not, we purge all the contents and register the packets as lost) becomes the new \textit{last} queue. The Paternoster operation repeats at each \textit{epoch}, while the four queues alternate during each \textit{epoch}.  While four queues are expected to be sufficient for many LDN scenarios, very long propagation delays may necessitate that another queue into the past and another queue into the future are added, for a total of six queues~\cite{seaman2019pat}.

In summary, the Paternoster approach uses four queues that alternate every epoch (also known as cycle) using only frequency synchronization, i.e., the epoch duration is the same across the nodes. In contrast to CQF, the Paternoster approach gives up some delay predictability in exchange for not requiring clock synchronization and for reducing the average delay.

The evaluations reported for Paternoster in this report considered
random time shifts of the equal-duration cycles in the switches.  In
particular, each switch had an independent uniformly distributed time
shift between zero and the cycle time with respect to a common time
base.

\subsection{3-Queue CQF}

\begin{figure} [t!]
	\centering
	\includegraphics[width=3.5in]{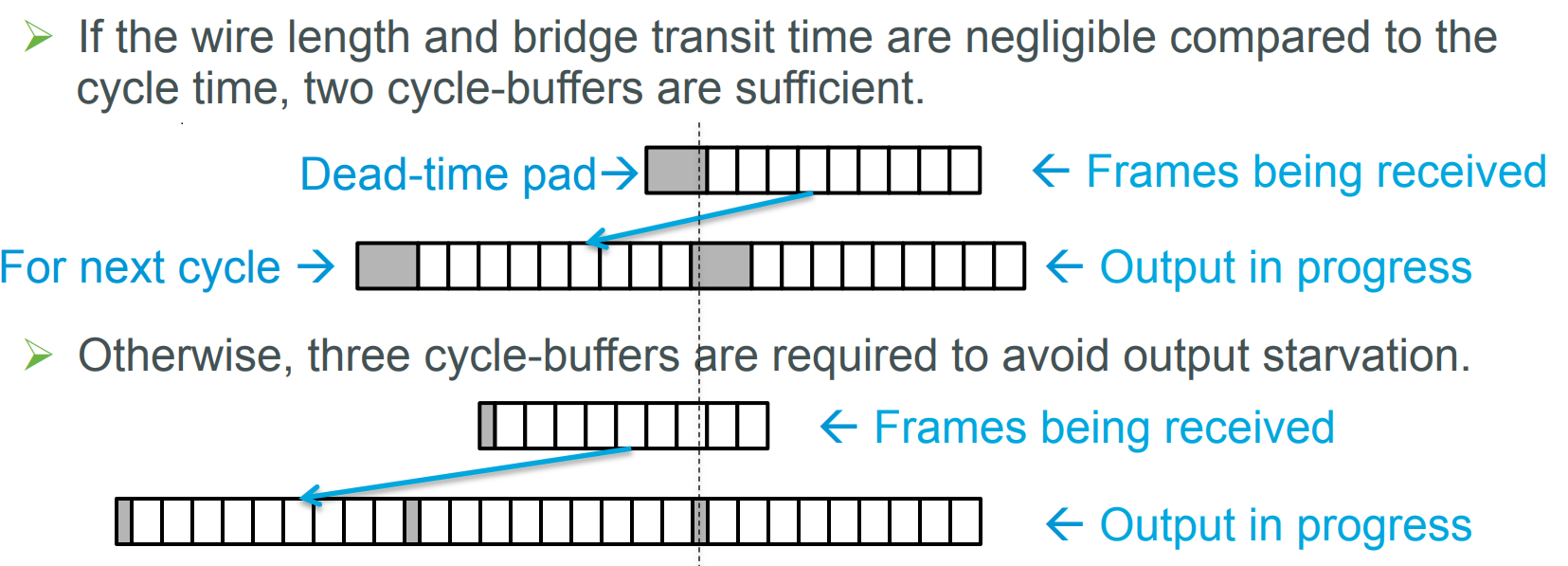}
	\caption{Standard vs. 3-Queue CQF example illustration.}
	\label{3Q-CQF}
\end{figure}

A critical requirement for the standard CQF is that a frame sent during a cycle has to be received during the same cycle such that the worst case delay is constrained by the cycle time and hop count. The 3-Queue CQF has been proposed to handle networks that have propagation delays that approach and exceed the cycle time~\cite{finn-detnet-bounded-latency-03}.

When traffic arrives in the wrong cycle, a $3^{rd}$ queue is needed to handle such traffic so as to prevent disruption to traffic that conforms to the requirement for CQF. This is illustrated in Fig.~\ref{3Q-CQF}. Though the general or principle idea behind the 3-Queue CQF is interesting, some questions remain that need solutions so that a full-fledged implementation and evaluation is possible.

How would the $3^{rd}$ queue (or waiting queue) be used in such a
environment without affecting other traffic? Every cycle is needed to
send traffic from an egress port, especially for periodic
traffic. Therefore, when should traffic that gets enqueued into the
waiting queue be dequeued? How would the dead-time be calculated or
computed? If the propagation delay exceeds the cycle time for all
periodic traffic, wouldn't this delay be consistent for all traffic and
therefore act as a constant in the overall worst case delay?

Our tests
in Section~\ref{tsn:sim:setup} indicate that a propagation delay of
$50\mu$s for a $50\mu$s cycle time (where ST is given $25\mu$s) gives
twice (from $200\mu$s to $400\mu$s) the max or worst case delay than a
propagation delay of $25\mu$s. We can hypothesize that for sporadic sources,
we can use the strict priority scheduler between the dequeuing queue
and the waiting queue, so that any traffic in the waiting queue can be
transmitted if no traffic is waiting in the dequeuing queue.

%
%
%
%

\section{Performance Evaluation}  \label{tsn:sim:setup}

\subsection{System Overview and Simulation Setup}  \label{tsn:eval:sec}
This section describes the simulation setup and model for both standard CQF and the Paternoster scheduling protocols. Furthermore, the topology and simulation scenarios will be presented. Throughout, we employ the OMNet++~\cite{varga2008overview} simulation environment.

\begin{figure} [t!]
	\centering
	\includegraphics[width=3.5in]{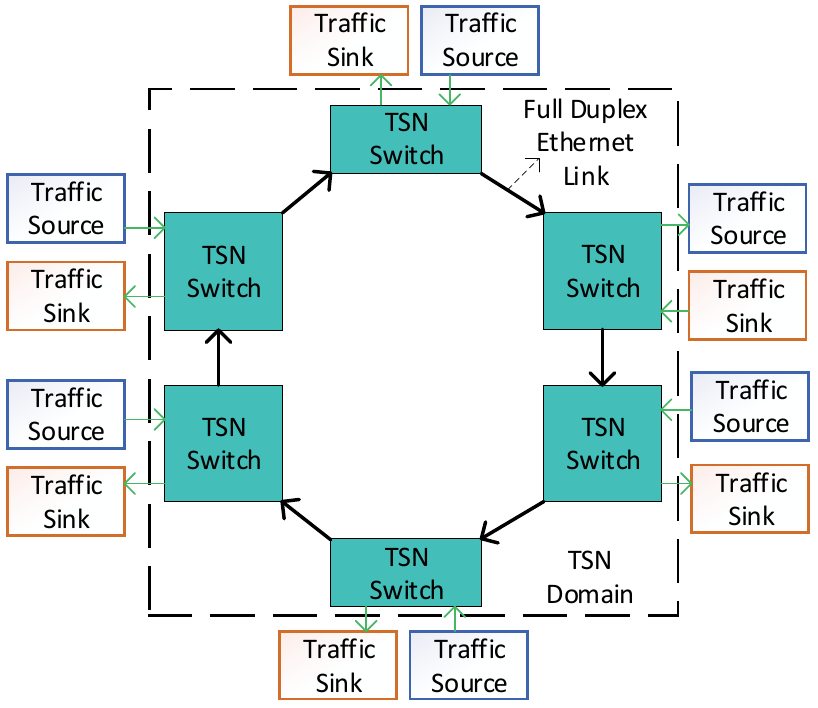}
	\caption{Unidirectional Ring Topology}
	\label{ring}
\end{figure}

\begin{table}[t!]
	\footnotesize
	\centering
	\caption{Simulation Parameters}
	\begin{tabularx}{\columnwidth}{|X|X|X|}
		\hline
		\multicolumn{1}{|c|}{\textbf{Key}} & \multicolumn{1}{|c|}{\textbf{Symbol}} & \multicolumn{1}{c|}{\textbf{Value}} \\
		\hline
		Simulation duration & $Sim_{limit}$ &  $100$ seconds \\
		\hline
		Initialized cycle time & $GCL_{CT}$ & $50~\mu$s \\
		\hline
		Initialized gating ratio & $ST^{R}_{init}$ & $50\%$ (i.e., 25~$\mu$s) \\
		\hline
		Total streams & $\gamma$ & $6$ \\
		\hline
		Stream duration & $\tau$ & $100$ seconds \\
		\hline
	Link propagation delay & $\alpha$ & $500~ns$, $25\mu$s,
                $50\mu$s \\
		\hline
		Number of frames/packets per cycle for periodic traffic & $\pi$ & $1 - 40$ \\
		\hline
		Sporadic traffic intensity & $\rho_{I}$ & $0.1 - 2.0$ Gbps \\
		\hline
		ST sources & $S$ & $6$ \\
		\hline
		ST stream hop count & $TTL$ & $3$ \\
		\hline
		Hurst parameter & $H$ & $0.5$ \\
		\hline
		Queue size & $Q_{size}$ & $512$~Kb \\
		\hline
	\end{tabularx}
	\label{table: simulation parameters}
\end{table}

\subsubsection{Network Model}	\label{tas:sec:net}
The topology used to test the CQF and Paternoster scheduling mechanisms is modeled as shown in Fig.~\ref{ring}. Table.~\ref{table: simulation parameters} shows the simulation parameters used in testing CQF and Paternoster in the unidirectional ring. Each switch-to-switch link operates as a full-duplex Ethernet link with a capacity (transmission bitrate) $R = 1$~Gbps. Each switch can act as a gateway for a number of traffic sources and one sink. The propagation distance is varied between $500$~ns and $50\mu$s. Each switch operates either CQF or Paternoster scheduling between switch to switch egress ports.

\subsubsection{Traffic Model}		\label{tas:sec:traffic}
We consider periodic (pre-planned) traffic and sporadic self-similar Poisson ($H = 0.5$) traffic for ST traffic, while solely sporadic traffic for BE.  Six sources are used to generate traffic each attached to a TSN switch gateway. A single stream is initiated at the start of the simulation for the entire duration of the simulation. Each frame/packet's destination address is specified by the switch to switch hops around the ring, which is predefined to $3$ hops as shown the Table.~\ref{table: simulation parameters}. The size of a frame is $64$~bytes for ST and $580$~bytes for BE. The traffic intensity is varied in each simulation run where the ST injection rate ($1-40$) is used for periodic ST traffic and the $\rho_{I}$ traffic intensity. Note that the BE traffic intensity in periodic ST source tests is set to $1.0$~Gbps.

\subsection{CQF}

\subsubsection{Periodic Results}

\begin{figure} [t!]
	\centering
	\includegraphics[width=3.5in]{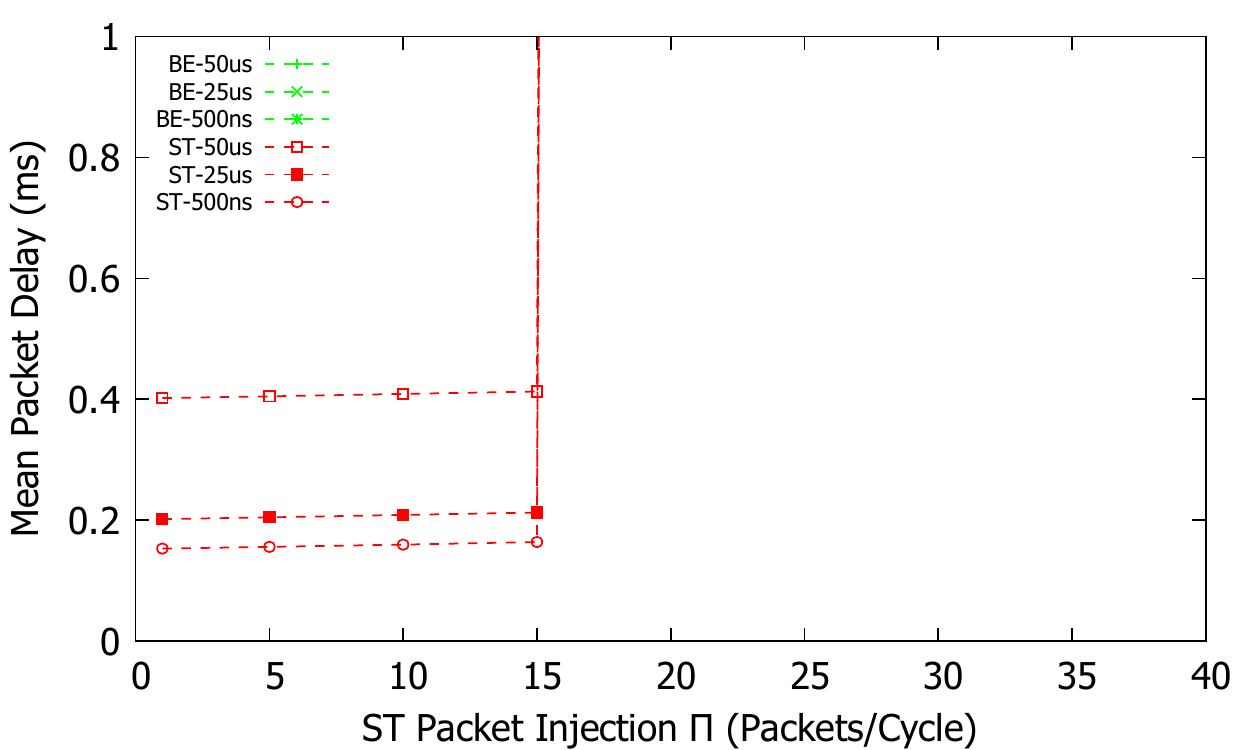}
	\caption{Periodic traffic ST sources .. CQF mean packet delay}
	\label{cqf-mean-periodic}
\end{figure}

\begin{figure} [t!]
	\centering
	\includegraphics[width=3.5in]{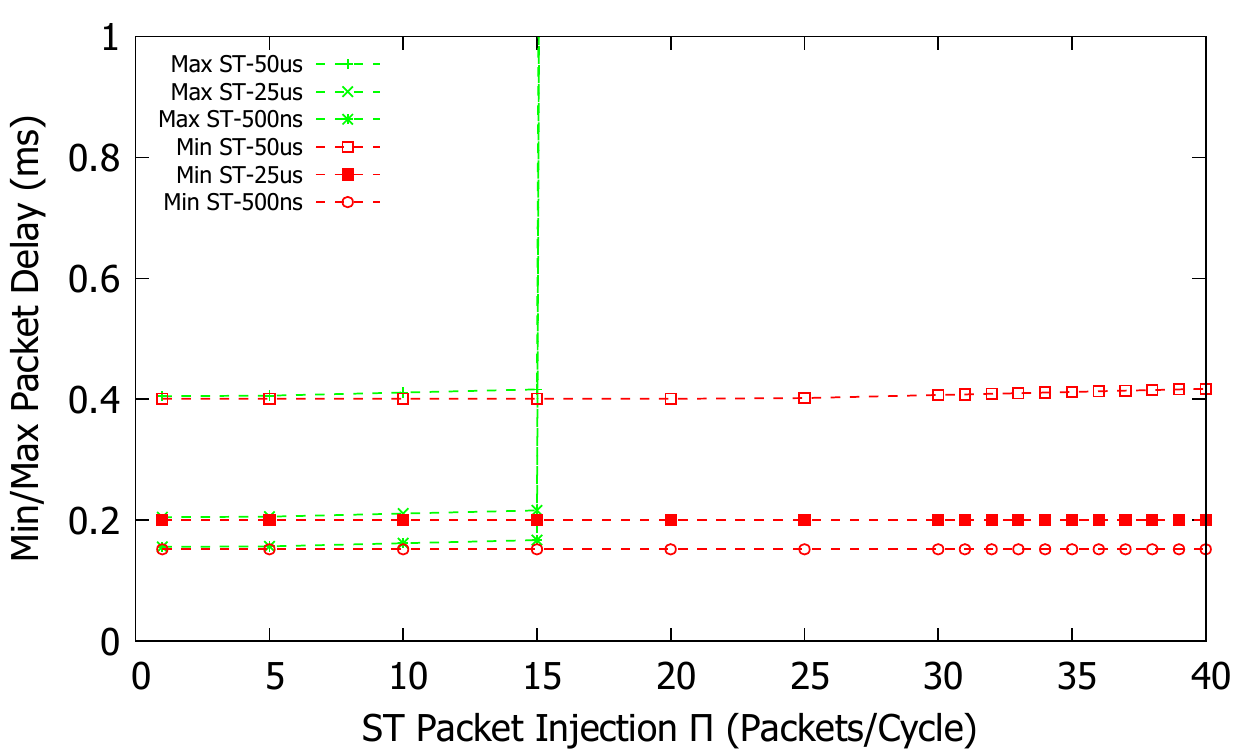}
	\caption{Periodic traffic ST sources .. CQF max/min packet delay}
	\label{cqf-max-periodic}
\end{figure}

Fig.~\ref{cqf-mean-periodic} and Fig.~\ref{cqf-max-periodic} show the
mean and min/max delays, respectively, for periodic ST traffic using
CQF based scheduling under different propagation delays. The BE
traffic intensity is set to a constant value of $1.0$~Gbps and
exhibits the same mean delay of $28$~ms for all ST injection rates
(due to TAS isolation). As the ST periodic traffic intensity
increases, both the mean and max delays are constant up to an ST
packet injection rate of $\pi = 16$ packets/cycle, which causes an
immediate spike in both mean and max delays due to over-utilizing the
link resources, no preventive measures of admission control policies,
and none-adaptive TAS slot ratios that change according to the
bandwidth consumption. Since CQF gives a simple method of calculating
the worst-case end-to-end delay of a stream (shown in
Section~\ref{tsn:back:sec}), the max delay shown in
Fig.~\ref{cqf-max-periodic} illustrates that for the CQF mechanism, the
delay is a function of and bounded by the number of hops and GCL
time. More precisely, since the cycle time (GCL) is set to $50\mu$s,
the total worst case delay for a three hop stream is $50 \cdot (3+1) = 
200\ \mu$s which is shown in both figures (except for networks
initialized with $50\mu$s propagation delays).

Furthermore, as the propagation delay is increased and approaches the cycle time, the end-to-end delay approaches the CQF worst-case delays, i.e., the min/max and mean delays are bounded and characterized by the number of hops and cycle time. Note that the ST gating ratio (due to TAS operating in the egress port) is half the cycle time ($25\mu$s), while the BE traffic is allocated the rest of the transmission time opportunity. When the propagation approaches the cycle time, it is considered twice the ST gating ratio which translates to twice the worst case delay of $400\mu$s.

\begin{figure} [t!]
	\centering
	\includegraphics[width=3.5in]{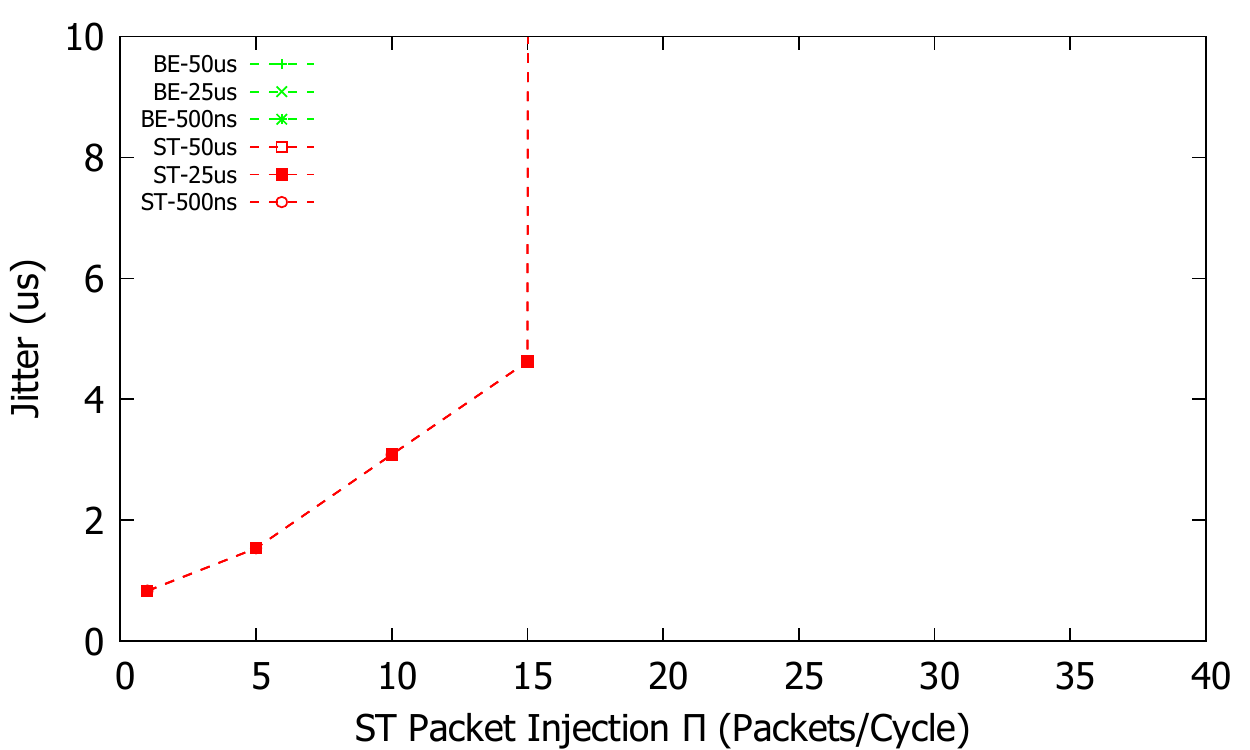}
	\caption{Periodic traffic ST sources .. CQF Jitter}
	\label{cqf-jitter-periodic}
\end{figure}

Fig.~\ref{cqf-jitter-periodic} shows the network jitter between a source and sink. The jitter is calculated as the standard deviation of the mean delay. As shown in the figure, jitter is around $4\mu$s but then spikes very quickly due to over-utilization of resources and consequently causing congestion.

\begin{figure} [t!]
	\centering
	\includegraphics[width=3.5in]{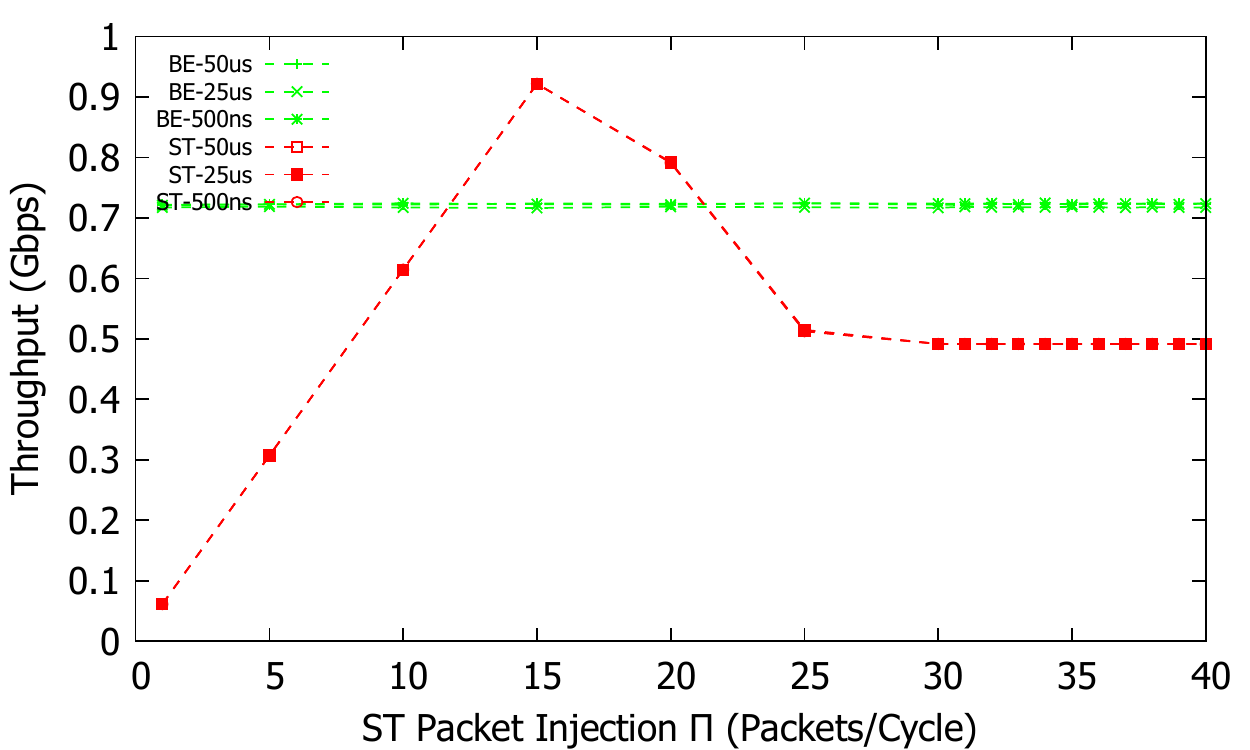}
	\caption{Periodic traffic ST sources .. CQF Average Throughput}
	\label{cqf-tput-periodic}
\end{figure}

\begin{figure} [t!]
	\centering
	\includegraphics[width=3.5in]{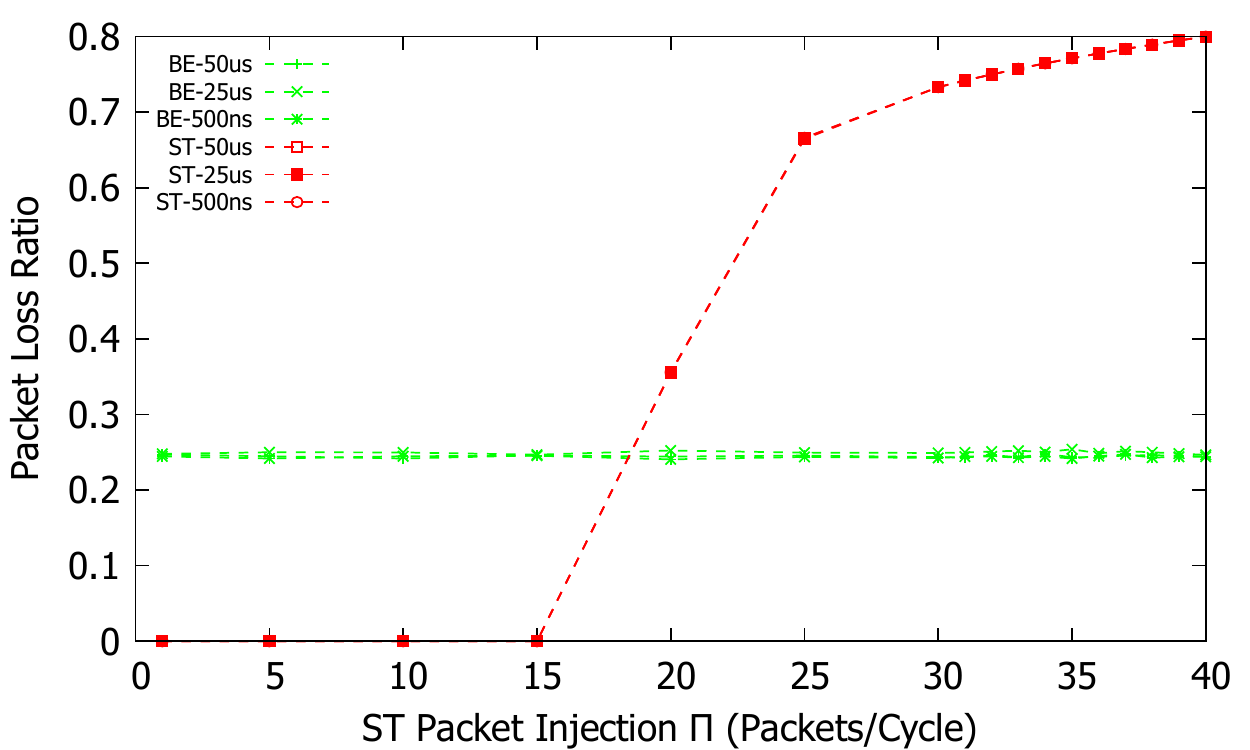}
	\caption{Periodic traffic ST sources .. CQF Loss Packet Ratio}
	\label{cqf-loss-periodic}
\end{figure}

Fig.~\ref{cqf-tput-periodic} and Fig.~\ref{cqf-loss-periodic} show
the throughput and loss respectively.
As the ST injection rate is increased, the throughput increases
linearly. However, the throughput sharply declines due to the
congestion caused by injecting more bits than the ST slot can handle
within each cycle ($16$ packets or $1892$ bits can be sent by one
source per cycle into the network).

In terms of packet/frame loss, BE traffic experiences more or less the same loss due to having the same traffic intensity ($1.0$~Gbps) in all runs (around $0.25\%$ loss). ST, on the other hand, stays constantly at $0$ loss until $\pi = 16$ rate. Due to congestion, the loss increases sharply with higher ST traffic intensity.

\subsubsection{Sporadic Results}

\begin{figure} [t!]
	\centering
	\includegraphics[width=3.5in]{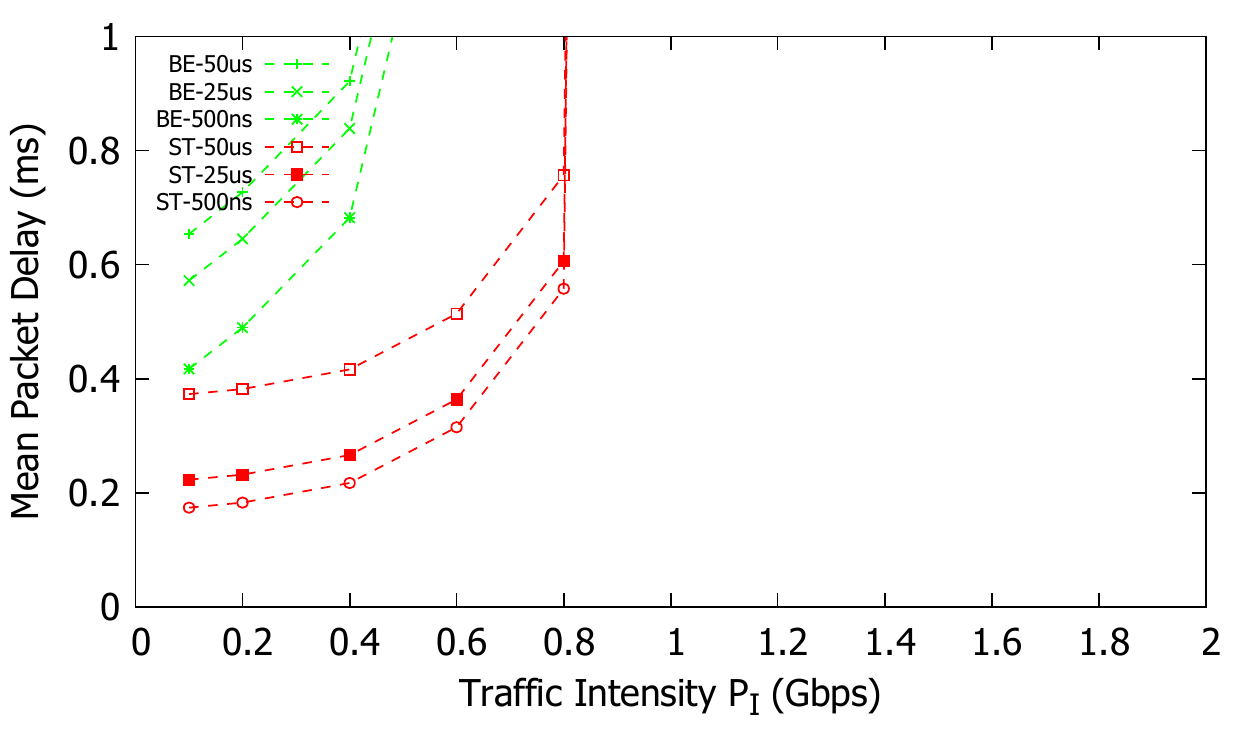}
	\caption{Sporadic traffic ST sources .. CQF mean packet delay}
	\label{cqf-mean-sporadic}
\end{figure}

\begin{figure} [t!]
	\centering
	\includegraphics[width=3.5in]{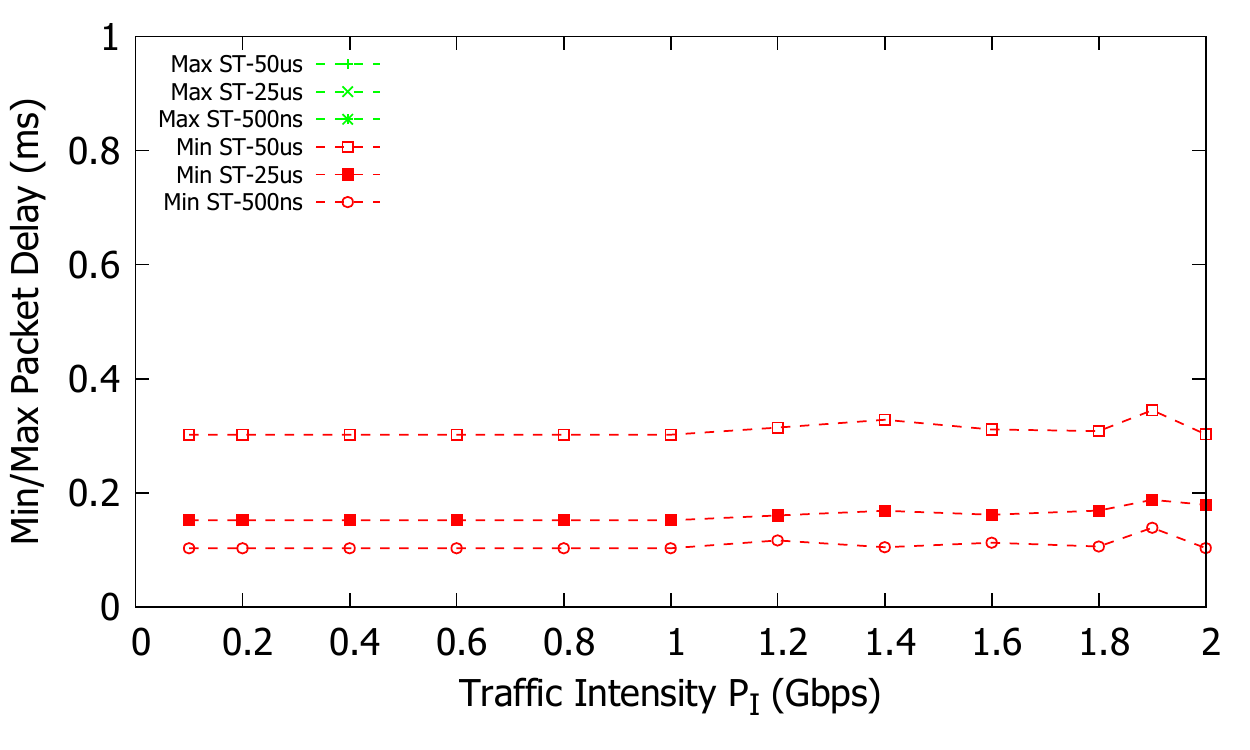}
	\caption{Sporadic traffic ST sources .. CQF max/min packet delay}
	\label{cqf-max-sporadic}
\end{figure}

Fig.~\ref{cqf-mean-sporadic} and Fig.~\ref{cqf-max-sporadic} show the mean and min/max delays respectively for sporadic ST sources. In contrast to periodic ST traffic sources, the use of sporadic traffic with uncontrollable bursts can severely degrade the operation of CQF as shown in both figures. The mean delay for both sporadic traffic classes quickly increases as the traffic intensity increases. The TSN QoS (bounded min/max delays, zero loss, and low jitter) are violated mainly due to the uncontrollable bursts in the sporadic ST sources.

\begin{figure} [t!]
	\centering
	\includegraphics[width=3.5in]{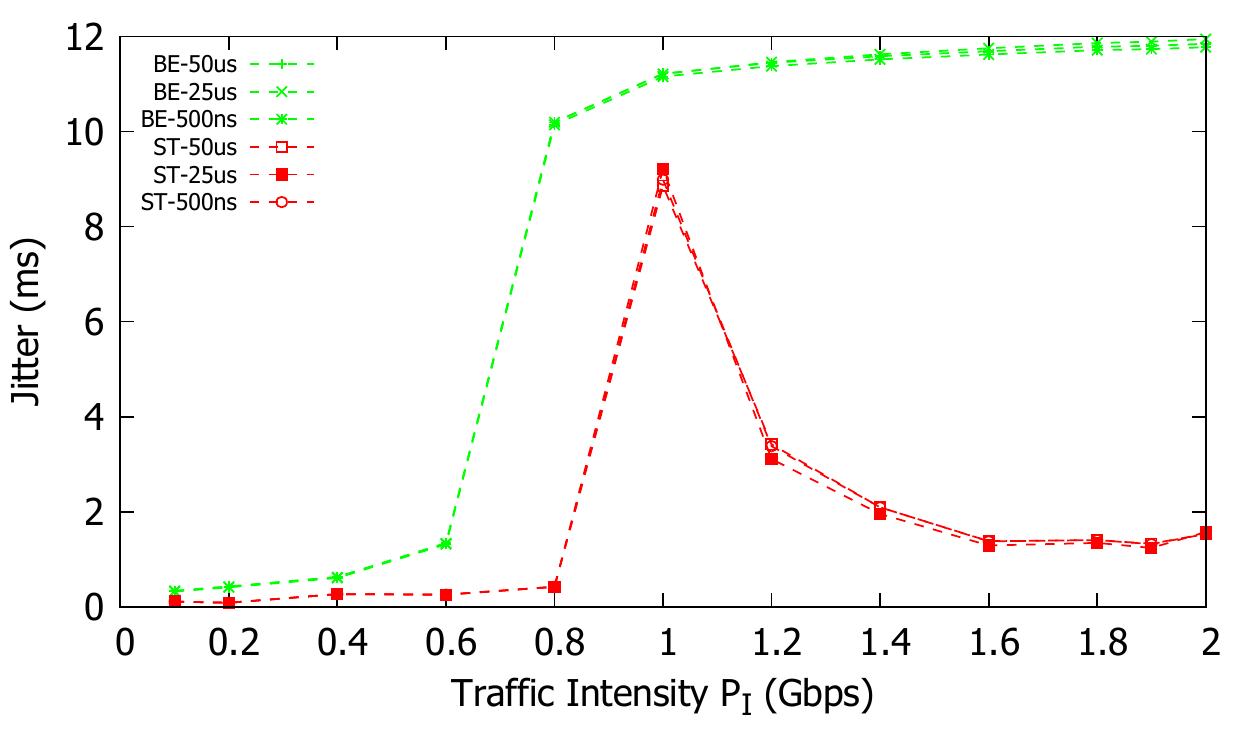}
	\caption{Sporadic traffic ST sources .. CQF Jitter}
	\label{cqf-jitter-sporadic}
\end{figure}

Fig.~\ref{cqf-jitter-sporadic} shows the network jitter between source and sink. Similar to the mean and min/max delays figures, the jitter is much higher compared to the periodic jitter results.

\begin{figure} [t!]
	\centering
	\includegraphics[width=3.5in]{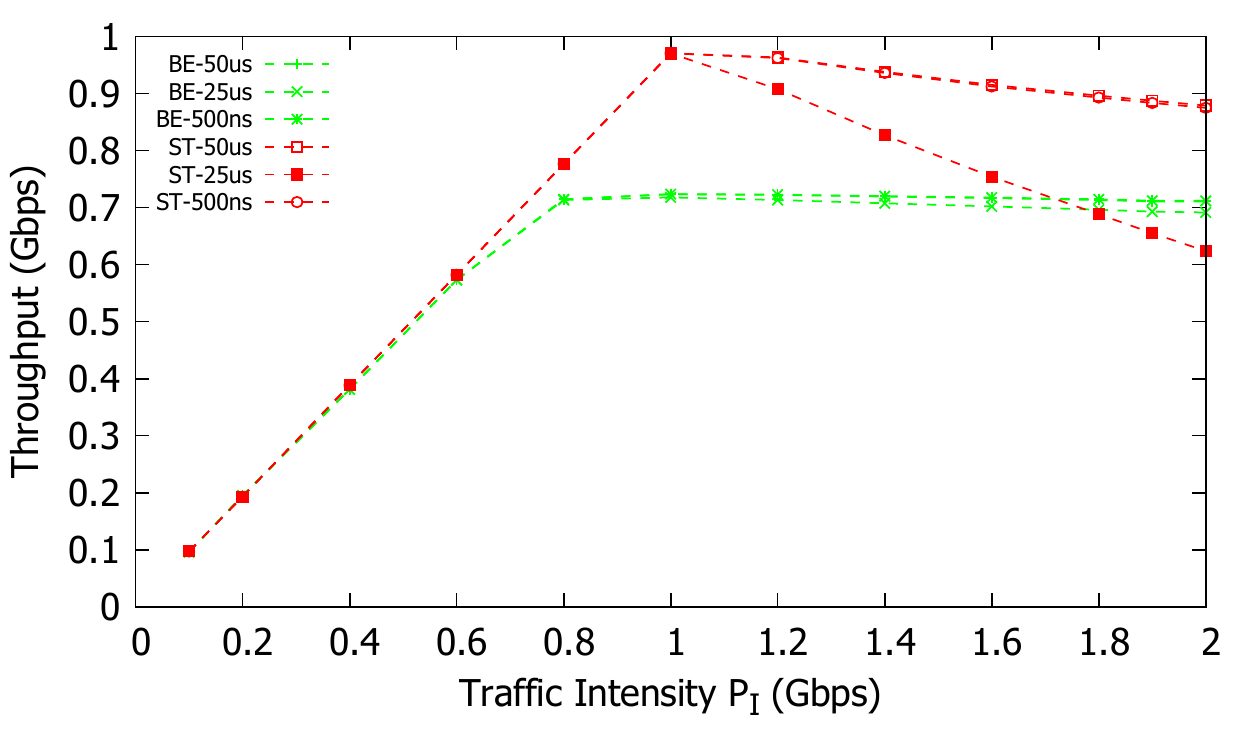}
	\caption{Sporadic traffic ST sources .. CQF Average Throughput}
	\label{cqf-tput-sporadic}
\end{figure}

\begin{figure} [t!]
	\centering
	\includegraphics[width=3.5in]{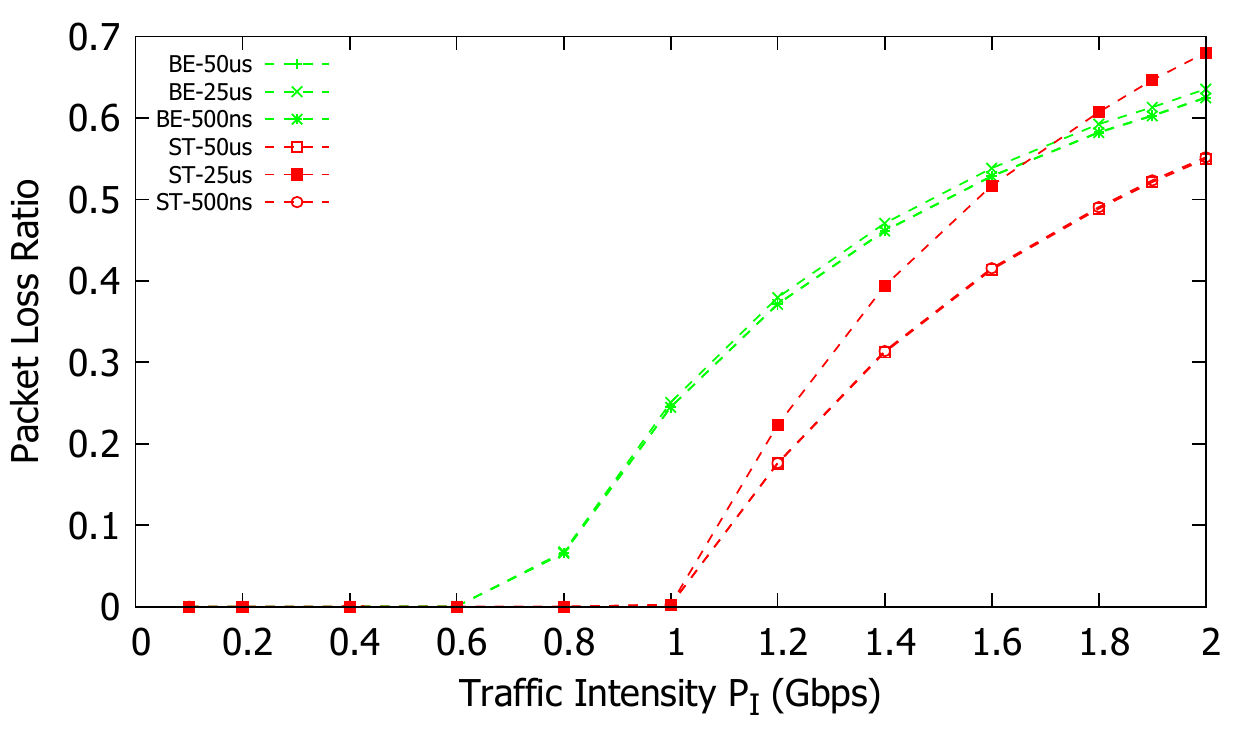}
	\caption{Sporadic traffic ST sources .. CQF Loss Packet Ratio}
	\label{cqf-loss-sporadic}
\end{figure}

Fig.~\ref{cqf-tput-sporadic} and Fig.~\ref{cqf-loss-sporadic} show the throughput and packet loss for sporadic ST sources. Throughput increases as the traffic intensity increases up to traffic intensity, $\rho_{I} = 1.0$, which causes a large drop in throughput due to congestion in the network. Similarly, the packet loss shows large increase after $\rho_{I} = 1.0$ for ST, while BE starts to lose more packets earlier.

\subsection{Paternoster}

\subsubsection{Periodic Results}

\begin{figure} [t!]
	\centering
	\includegraphics[width=3.5in]{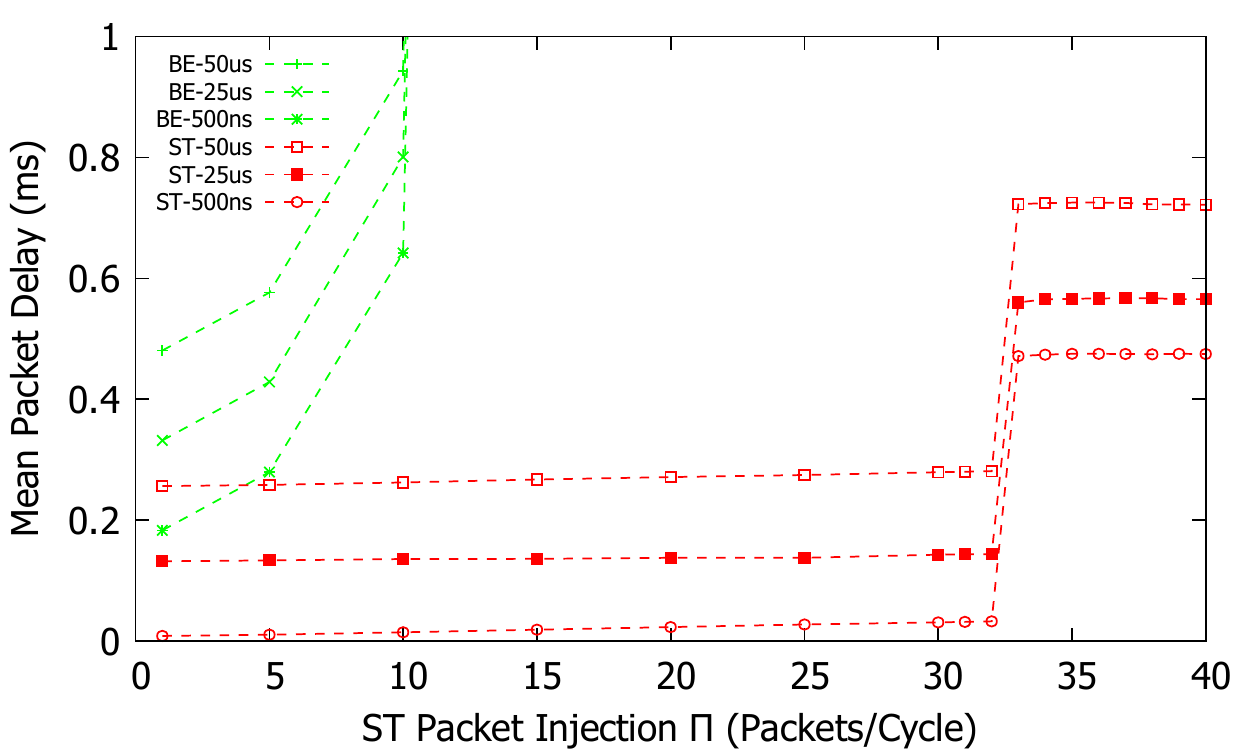}
	\caption{Periodic traffic ST sources .. Paternoster mean packet delay}
	\label{pat-mean-periodic}
\end{figure}

\begin{figure} [t!]
	\centering
	\includegraphics[width=3.5in]{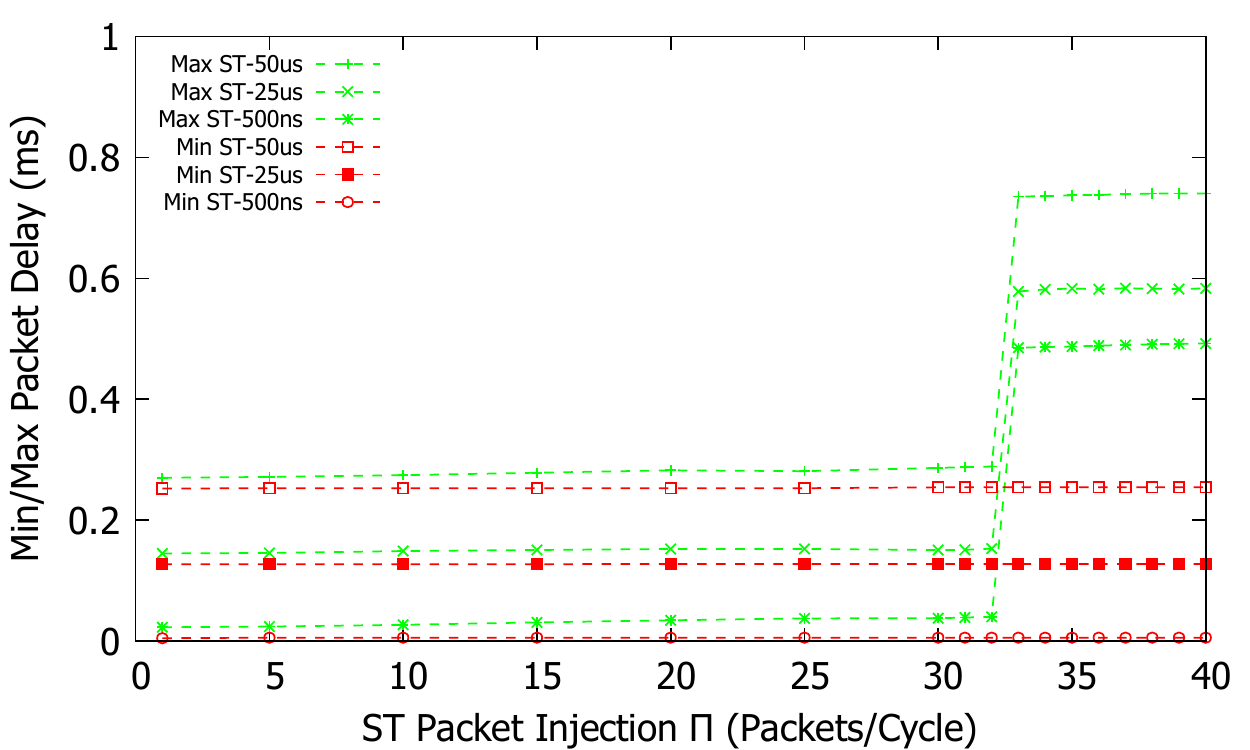}
	\caption{Periodic traffic ST sources .. Paternoster max/min packet delay}
	\label{pat-max-periodic}
\end{figure}

Fig.~\ref{pat-mean-periodic} and Fig.~\ref{pat-max-periodic} show the mean and max/min delays for periodic ST sources and sporadic BE sources using switches that operate Paternoster. Initially, we observe from Fig.~\ref{pat-mean-periodic} that the mean delays for ST are lower when compared against the CQF performance. However, BE get starved by ST when $\pi = 33$ since all transmission opportunities during an epoch/cycle are consumed by ST. ST's delay stabilizes after the spike due to purging the prior queue in Paternoster.

\begin{figure} [t!]
	\centering
	\includegraphics[width=3.5in]{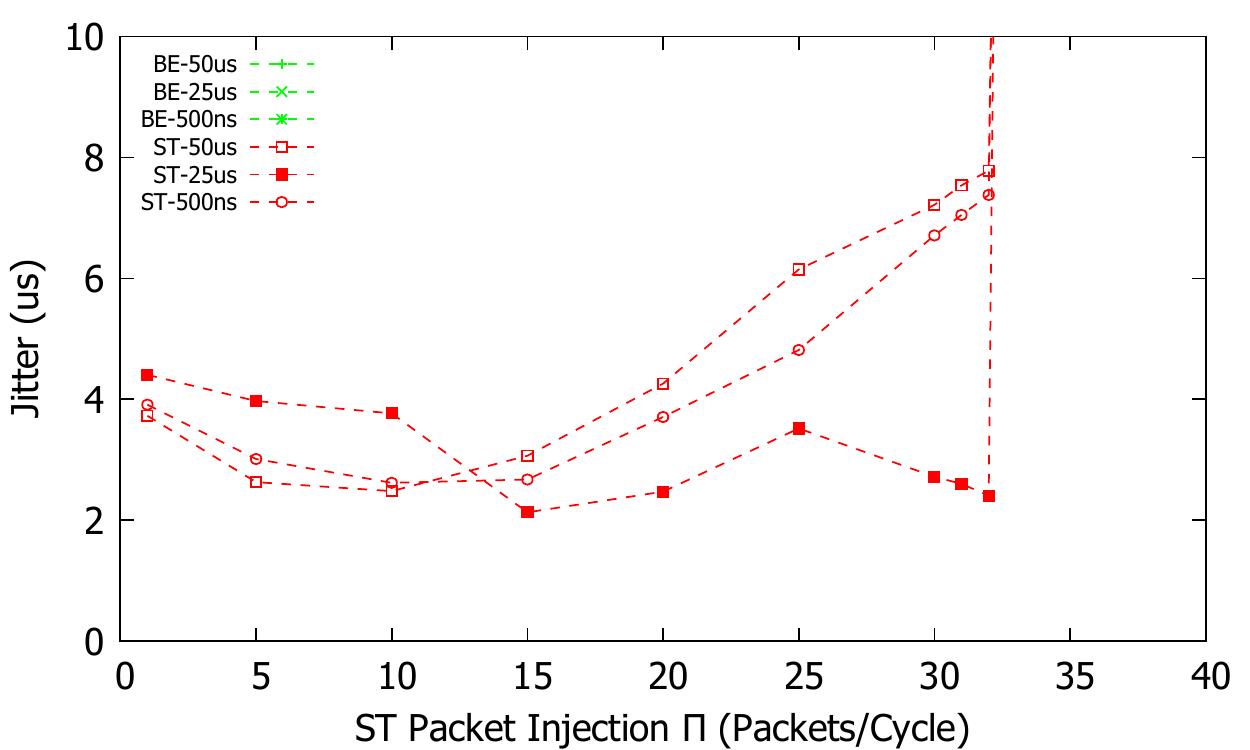}
	\caption{Periodic traffic ST sources .. Paternoster Jitter}
	\label{pat-jitter-periodic}
\end{figure}

Fig.~\ref{pat-jitter-periodic} shows network jitter. While the jitter is comparable to the CQF protocol, the varying changes as the traffic intensity increases show the unpredictability in Paternoster compared to CQF.

\begin{figure} [t!]
	\centering
	\includegraphics[width=3.5in]{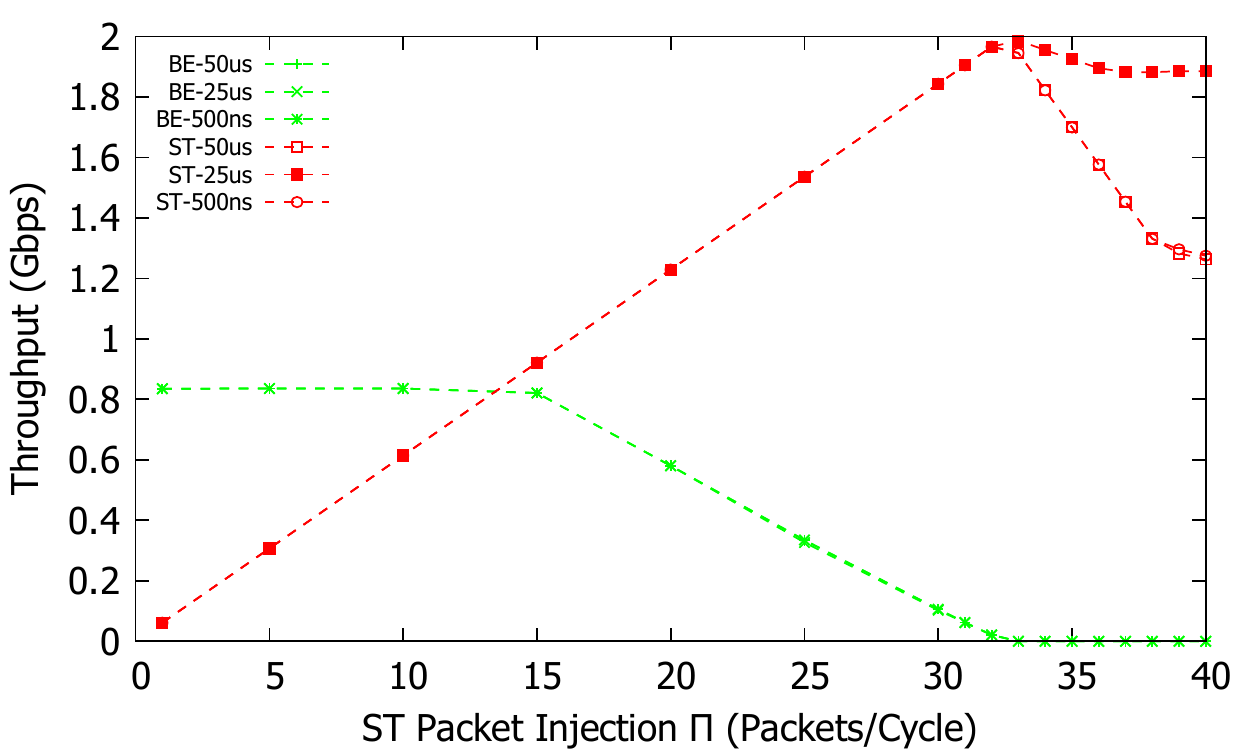}
	\caption{Periodic traffic ST sources .. Paternoster Average Throughput}
	\label{pat-tput-periodic}
\end{figure}

\begin{figure} [t!]
	\centering
	\includegraphics[width=3.5in]{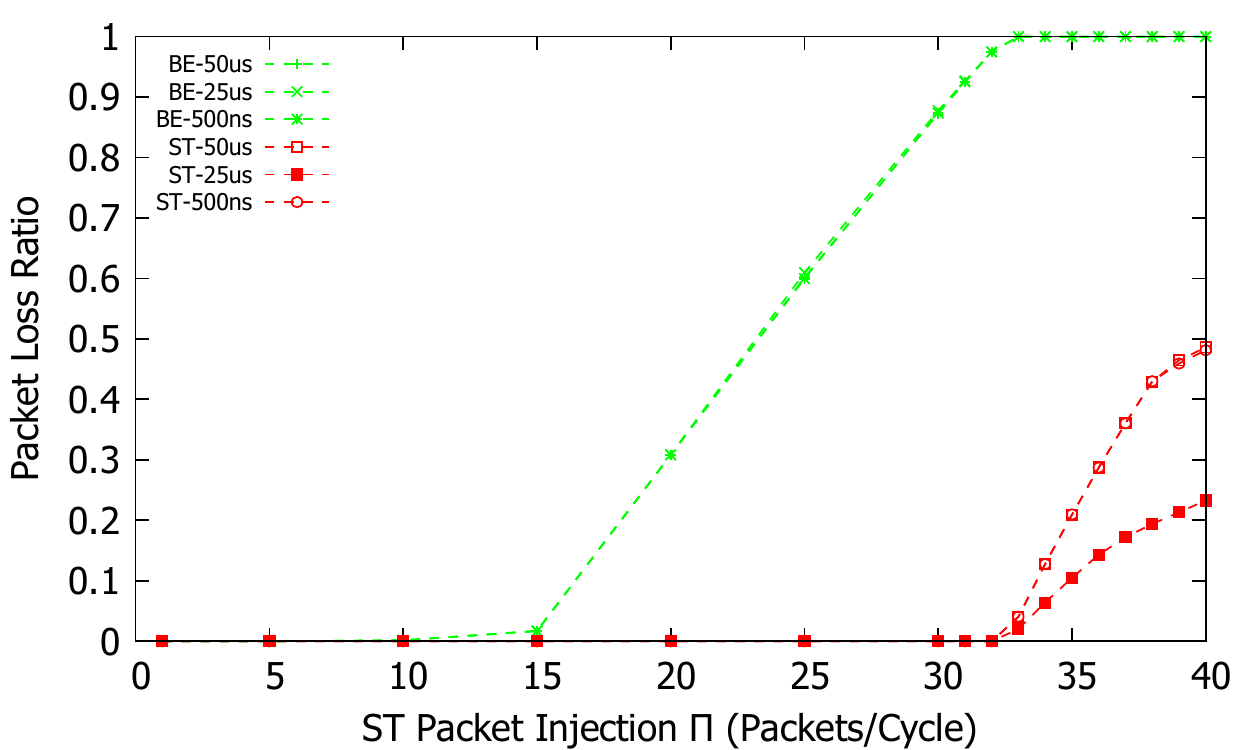}
	\caption{Periodic traffic ST sources .. Paternoster Loss Packet Ratio}
	\label{pat-loss-periodic}
\end{figure}

Fig.~\ref{pat-tput-periodic} and Fig.~\ref{pat-loss-periodic} show the throughput and packet loss experienced at the sink. At $\pi = 16$, we see the BE traffic (with $\rho_{I} = 1.0$~Gbps) starts to drop proportional to the increase in ST before being starved at $\pi = 33$. Any additional increase causes packet loss and congestion which drops the throughput to below optimum levels. Similarly, the loss shows a complement of the throughput and at $\pi = 16$, the BE traffic starts to accumulate loss linearly as the traffic intensity keeps increasing.

\subsubsection{Sporadic Result}

\begin{figure} [t!]
	\centering
	\includegraphics[width=3.5in]{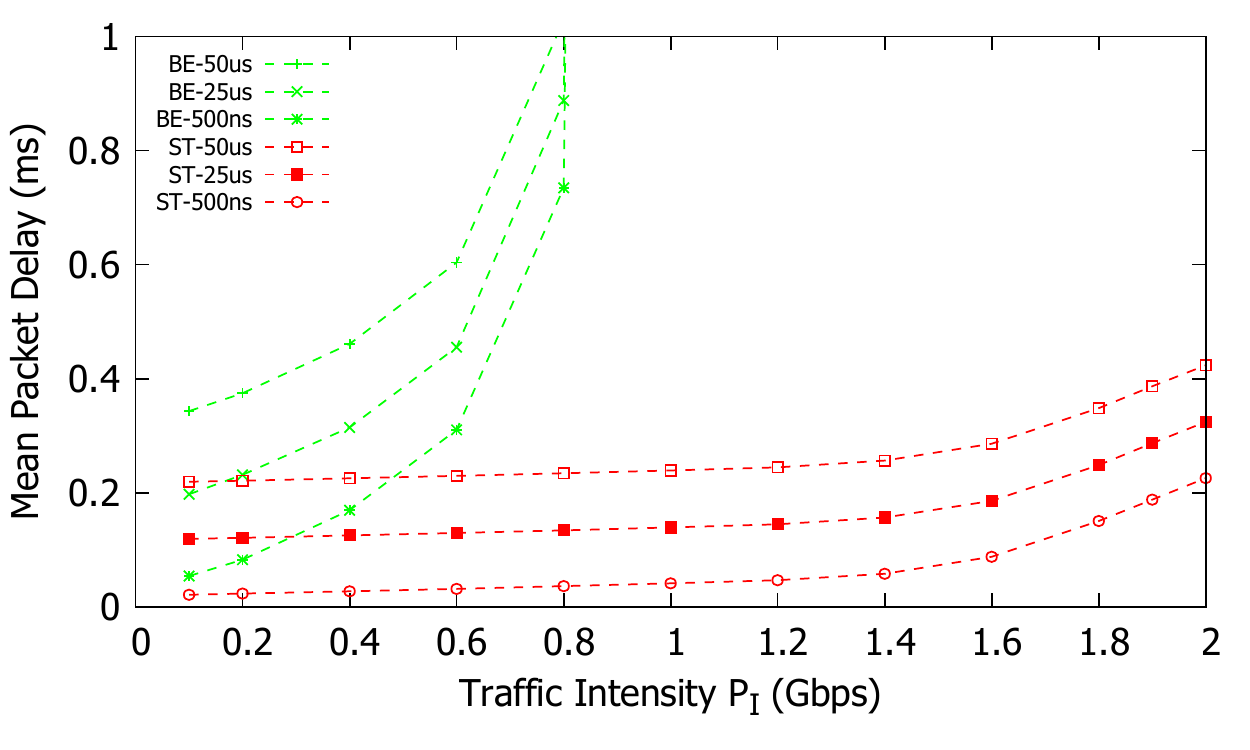}
	\caption{Sporadic traffic ST sources .. Paternoster mean packet delay}
	\label{pat-mean-sporadic}
\end{figure}

\begin{figure} [t!]
	\centering
	\includegraphics[width=3.5in]{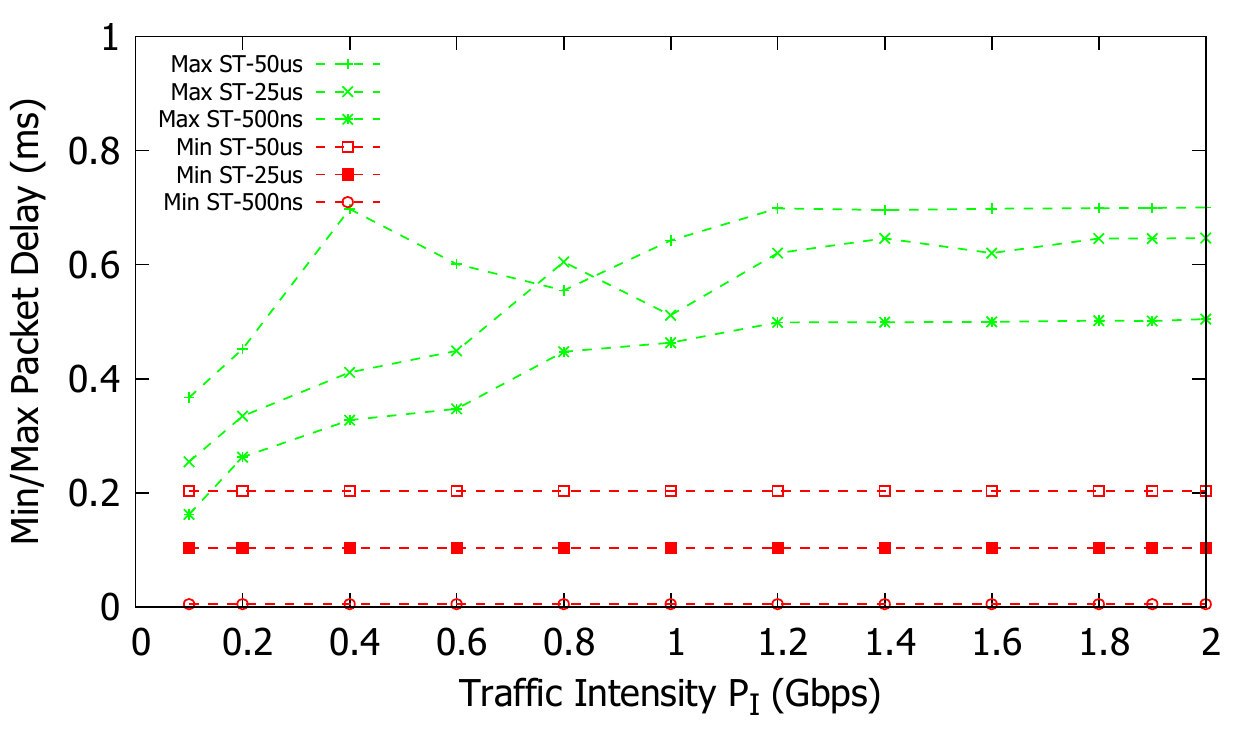}
	\caption{Sporadic traffic ST sources .. Paternoster max/min packet delay}
	\label{pat-max-sporadic}
\end{figure}

Fig.~\ref{pat-mean-sporadic} and Fig.~\ref{pat-max-sporadic} show the mean and min/max delays for sporadic ST sources using Paternoster. Compared to the periodic results, the Paternoster performs better for sporadic traffic sources. However, accurately predicting the worst case delays still remains difficult compared to CQF. Since strict priority scheduling is employed at the egress port, the max or worst case delays are highly unpredictable compared to CQF and also Paternoster under the periodic ST sources.

\begin{figure} [t!]
	\centering
	\includegraphics[width=3.5in]{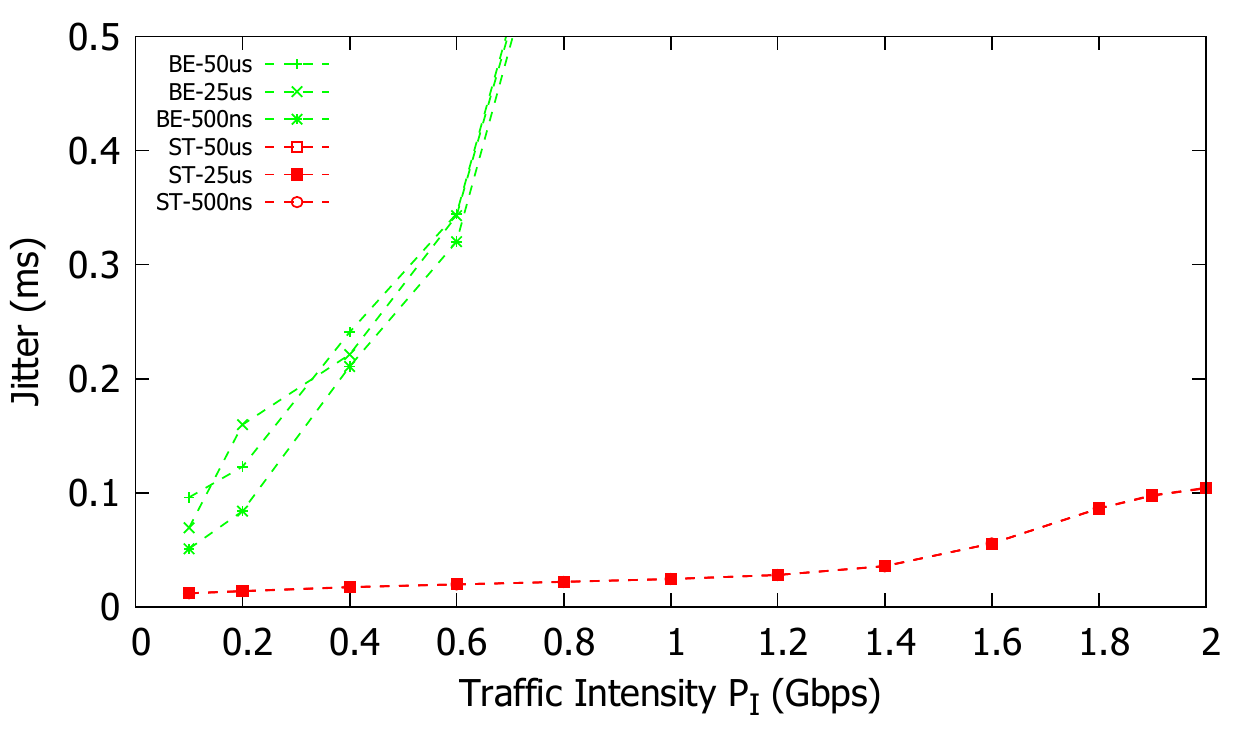}
	\caption{Sporadic traffic ST sources .. Paternoster Jitter}
	\label{pat-jitter-sporadic}
\end{figure}

Fig.~\ref{pat-jitter-sporadic} shows the network jitter. Since strict priority scheduling is used to arbitrate between competing traffic classes, BE (lower priority) can block ST (higher priority) if the port is currently transmitting BE traffic when ST waits for the transmission to finish. This causes higher unpredictable jitter, which is seen Fig.~\ref{pat-jitter-sporadic}.

\begin{figure} [t!]
	\centering
	\includegraphics[width=3.5in]{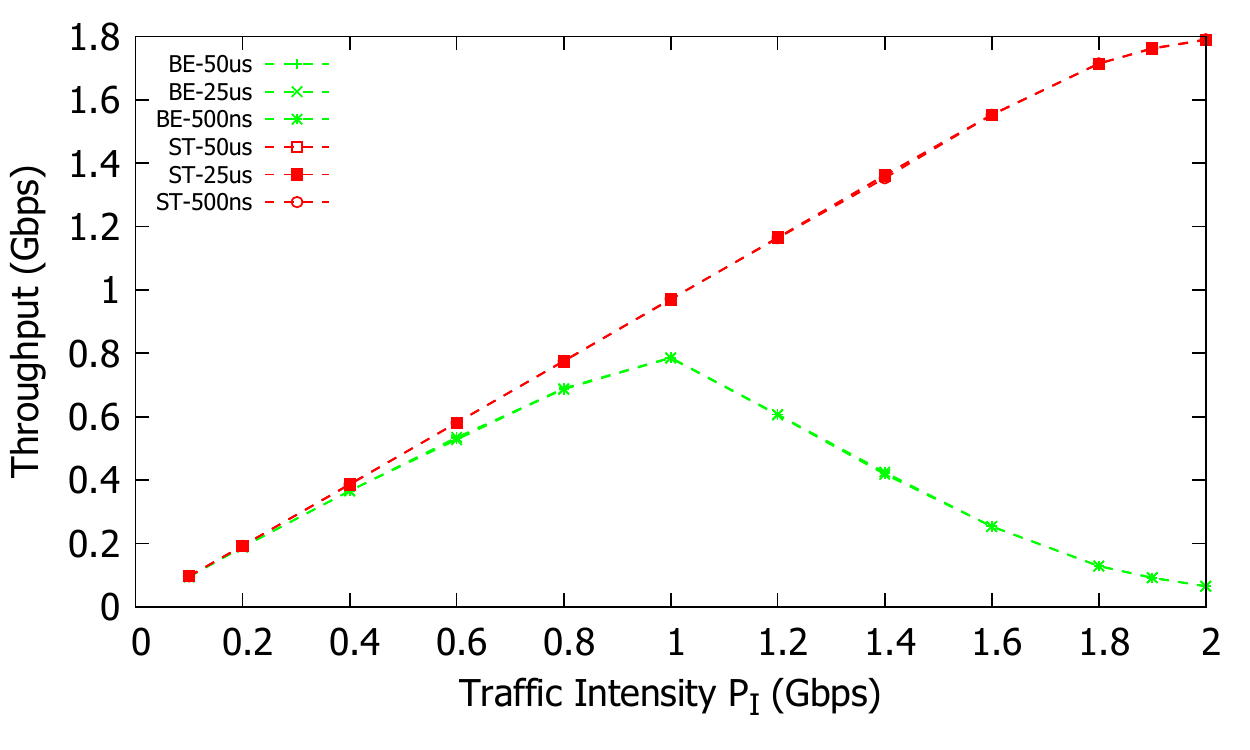}
	\caption{Sporadic traffic ST sources .. Paternoster Average Throughput}
	\label{pat-tput-sporadic}
\end{figure}

\begin{figure} [t!]
	\centering
	\includegraphics[width=3.5in]{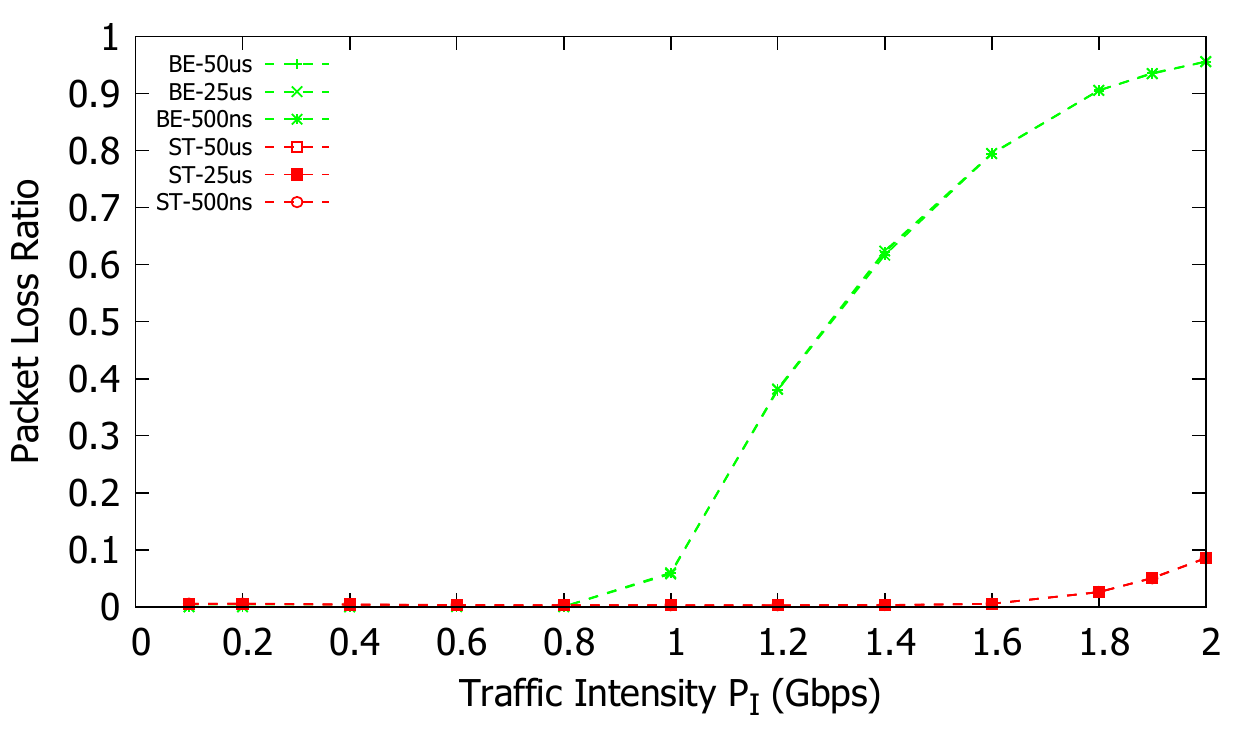}
	\caption{Sporadic traffic ST sources .. Paternoster Loss Packet Ratio}
	\label{pat-loss-sporadic}
\end{figure}

Fig.~\ref{pat-tput-sporadic} and Fig.~\ref{pat-loss-sporadic} show the
throughput and loss at the sink. Similar to the periodic case, BE
traffic throughput decreases after $1.0$~Gbps while it's loss
increases. In contrast, the ST throughput continues to increase, but
ST does experience some loss after about an injection rate $\pi$
corresponding to $1.6~Gbps$.

\subsection{CQF and Paternoster Comparison Analysis}
Both the CQF and Paternoster scheduling algorithms attempt to enhance the
data-link layer with some form of deterministic behavior. CQF
coordinates the ingress and egress operations to provide bounded
delays and zero traffic loss that conforms to its reservation, while
Paternoster uses four queues per port to spread traffic
burstiness/bunching and guarantees bounded delays for traffic that
conforms to its resource reservation. In essence, Paternoster applies
the concept of Credit-Based scheduling (CBS)~\cite{IEEE8021Q} with the
cyclic attribute of CQF.

Paternoster is considered an enhancement to CQF since it guarantees
TSN QoS whilst removing the time synchronization requirement, i.e., an
asynchronous scheduling protocol (though frequency synchronization is
still needed to keep the cycle/epoch duration the same at all
switches). While the results are favorable towards Paternoster in
terms of minimizing mean and max delay, CQF is generally more
predictable and therefore more deterministic than Paternoster,
particularly for OT applications with critical QoS and hard deadline
requirements. In particular, for periodic traffic and using the CQF
protocol (with $50\%$ gating ratio for ST, or $25\mu$s transmission
opportunity), all the streams with $\pi \le 16$ have mean/max/min
delays between $150\mu$s and $200\mu$s regardless of the path and
cross traffic. This bounded delay guarantee is not easily predictable
for Paternoster due to the fundamental loss of time synchronization
between switches. Additionally, since Paternoster uses the strict
priority scheduling at the egress, it contains elements of best-effort
service which causes loss of determinism.

CQF provides complete isolation between ST and BE (due to TAS being
used at the egress port) and as a result performs fairer in resource
allocation between ST And BE, especially at high traffic
intensity. Paternoster does not isolate traffic classes (though it
does provide resource allocation) which can degrade the predictability
and deterministic behavior for ST. This effect is evident in the
periodic jitter results where the mean jitter at varying traffic
intensities for CQF is monotonically increasing (up to a bounded
jitter value), while the jitter measurement for Paternoster is highly
erratic.

In terms of packet loss, both CQF and Paternoster guarantee zero loss
for streams that conform to their reservations. However, Paternoster
does perform slightly worse when packets remain in the \textit{prior}
queue and a cycle change occurs causing the \textit{prior} queue to be
purged of its content. This rarely happens with periodic traffic
sources, but can occur more frequently with sporadic traffic sources
where a switch can abruptly receive a large number of traffic before a
cycle change.

Since Paternoster uses strict priority scheduling at the egress,
Paternoster achieves significantly better delays due to having more
transmission opportunities than the CQF protocol, i.e., CQF's use of
TAS at the egress divides and isolates the transmission opportunities
and does not adapt these opportunities to varying changes in traffic
intensity. Moreover, the main issue with CQF in guaranteeing QoS for
sporadic ST streams is that the burst usually is much greater than the
allowable bandwidth per cycle (the transmission opportunities
given). Applying ingress policing and admission control (either
centralized or distributed) can mitigate this issue by using control
signals and negotiating network resources and QoS to streams that
request it (this has been investigated in~\cite{nas2019rec} switches
utilizing TAS only).

\section{Conclusions and Future Work}  \label{concl:sec}
A performance evaluation has been conducted in this report to compare
standard CQF and Paternoster. Since Paternoster uses more queues,
i.e., more complexity, and provides less deterministic behavior
compared to CQF, CQF performs better for OT applications with hard
real-time requirements. While Paternoster performs worse than CQF in
ensuring deterministic properties, it provides a relaxed traffic
predictability in networks that do not have time
synchronization. While this performance evaluation used statically
defined traffic slots in the cycle (for CQF/TAS), an adaptive method
(Adaptive TAS~\cite{nas2019per}) can be used to accurately determine
the needed slot duration to optimally service registered traffic
classes (in this case, BE and ST). Using TAS in Paternoster at the
egress port between ST and BE instead of strict priority scheduling is
another recommendation that can reduce the jitter for streams. While
the tests in this evaluation involved uniform link transmission and
propagation delays, a more complex problem that involves different
link transmission and propagation delays is an interesting direction
for future research since it can cause cycle misalignment between
adjacent ports according to the standard.

In the wider context of QoS networking and related applications,
QoS oriented routing approaches, e.g.~\cite{chu2018pre,guc2017uni}
should be investigated. Furthermore, 
deterministic networking should be studied in the context of
emerging multiple-access edge computing
(MEC)~\cite{doa2019pro,gao2019dyn,mar2019mod,sha2018lay,wan2019mul},
in particular MEC settings for low-latency
applications~\cite{elb2018tow,xia2019red,zha2018mob}.
As an alternative approach to coordinating the
reconfigurations, emerging softwarized control paradigms, such as
software defined networking can be explored~\cite{ami2018sdn,der2019cou,des2018min,kel2019ada,san2018inf}.
Regarding the reliability aspects, a potential future research direction is to
explore low-latency network coding mechanisms, e.g.,~\cite{ace2018har,coh2019ada,eng2018exp,gab2018cat,luc2018ful,ma2019high,wun2017cat}, to enhance networking protocols targeting reliable low-latency communication.

\bibliographystyle{IEEEtran}


\end{document}